\documentclass[prb,twocolumn,showpacs,preprintnumbers,amsmath,amssymb]{revtex4}
\usepackage{graphicx}
\usepackage{epsfig}
\begin{document}
\title{Non-linear theory of deformable superconductors}
\author{Pavel Lipavsk\'y$^{1,2}$, Klaus Morawetz$^{3,4}$, 
Jan Kol\'a\v cek$^2$ and Ernst Helmut Brandt$^5$}
\affiliation{
$^1$ Faculty of Mathematics and Physics, Charles University, 
Ke Karlovu 3, 12116 Prague 2, Czech Republic}
\affiliation{
$^2$Institute of Physics, Academy of Sciences, 
Cukrovarnick\'a 10, 16253 Prague 6, Czech Republic}
\affiliation{
$^3$Forschungszentrum Rossendorf, PF 51 01 19, 01314 Dresden, Germany}
\affiliation{
$^4$Max-Planck-Institute for the Physics of Complex
Systems, Noethnitzer Str. 38, 01187 Dresden, Germany}
\affiliation{
$^5$Max-Planck-Institute for Metals Research,
         D-70506 Stuttgart, Germany}
\begin{abstract}
The interaction of the superconducting condensate with deformations 
of the crystal lattice is formulated assuming the electrostatic 
potential to be of Bernoulli type and the effect of strain on 
material parameters. In the isotropic approximation it is shown 
that within the Ginzburg-Landau theory both contributions can be 
recast into the local but non-linear interaction term of the 
free energy.
\end{abstract}

\maketitle
\section{Introduction}
When cooled, metals reduce their volume. At the transition from normal 
to superconducting state, the coefficient of the thermal expansion 
makes a jump. In most cases, the superconducting systems reduce their 
volume less than the normal ones. Consequently, inhomogeneities of the 
superconducting phase cause stresses which are similar to stresses 
caused by the inhomogeneities due to the temperature.

Since the superconducting condensate affects the specific volume,
deformations of the crystal lattice also affect the condensate. 
Mechanisms of this interaction between the crystal lattice and the 
condensate can be outlined within the simplest two-fluid free energy 
of Gorter and Casimir, $f_{\rm GC}=-{1\over 4}\gamma T_{\rm c}^2\omega-
{1\over 2}\gamma T^2\sqrt{1-\omega}$, where $\omega$ is the 
superconducting fraction, $\gamma$ is the linear coefficient of the 
specific heat per unit volume known as the Sommerfeld $\gamma$, 
and $T_{\rm c}$ is the critical temperature. 

Material parameters $\gamma$ and $T_{\rm c}$ are not constants. They
depend on the electron density $n$ and the deformation of the crystal 
lattice which we describe by the lattice density $n_{\rm lat}$.
Due to the dependencies $\gamma(n,n_{\rm lat})$ and $T_{\rm c}(n,n_{\rm lat})$
the crystal lattice interacts with the superconducting condensate. 

In literature on deformable superconductors one \mbox{finds} two models 
of the lattice-condensate (lc) interaction. First, there are various
phenomenological theories\cite{Sc91,DuS92,C94,KBMD95,MDK95,CLM03} 
which assume that the density of electrons exactly follows the density 
of the lattice, $n=n_{\rm lat}$. The perturbation of material parameters
due to the deformation, e.g., $\delta\gamma=(\partial\gamma/\partial n)
\delta n+(\partial\gamma/\partial n_{\rm lat})\delta n_{\rm lat}$, then 
can be expressed via the lattice density, $\delta\gamma=
\left[(\partial\gamma/\partial n)+
(\partial\gamma/\partial n_{\rm lat})\right]\delta n_{\rm lat}$.
The strength of the lc-interaction 
thus depends on the sum of both density derivatives, 
$(\partial\gamma/\partial n)+(\partial\gamma/\partial n_{\rm lat})$ and 
$(\partial T_{\rm c}/\partial n)+(\partial T_{\rm c}/\partial 
n_{\rm lat})$. 

Second, a model in which the lc-interaction is mediated 
by the electrostatic potential has been discussed.\cite{Zhou94,LMKB07} 
Since the theory of the electrostatic potential has been developed 
under the approximation of a stiff lattice, the lc-interaction obtained 
within this model depends exclusively on the derivatives with respect 
to the electron density, $\partial\gamma/\partial n$ and $\partial 
T_{\rm c}/\partial n$.

The phenomenological approach is more general being 
applicable to all materials while the electrostatic approach is 
limited to cases in which the dependence on the electron density 
dominates. On the other hand, the electrostatic approach offers 
a natural picture of the surface, in particular, one can easily 
see that the electrostatic field of the surface dipole contributes 
to the forces deforming the crystal lattice.\cite{LMKBS08} Studies
within the phenomenological approach have not noticed the surface 
tension.

In this paper we derive a phenomenological theory which unifies both
approaches. To this end it is necessary to take into account the charge 
of a deformed lattice in the electrostatic potential and to allow for
lc-interaction which is not covered by the (mean) electrostatic 
potential. To avoid lengthy formulae or non-transparent tensor
notation with numerous indices, we restrict our attention to the
interaction between the lattice compression and the condensate. We
neglect the interaction between the condensate anisotropy and shear 
deformations of the lattice. Interaction of the superconducting
condensate with a deformation which can be interpreted as a mutual 
displacement of sublattices has been discussed in 
Ref.~[\onlinecite{SF89}].

\subsection{Origin of two mechanisms}

There are various microscopic mechanisms due to which material 
parameters of the superconductor depend either on the density of 
electrons or on the deformation of the crystal lattice. Although
our discussion is independent of actual microscopic mechanisms,
we find it profitable to outline some of these possibilities so that 
the need to treat perturbations of the electron density and the 
lattice density independently becomes more apparent.

The Sommerfeld $\gamma$ is proportional to the single-spin density of
states $N_0$ at the Fermi energy, $\gamma={2\over 3}\pi^2k_{\rm B}^2
N_0$. The density of states naturally depends on the value of the Fermi 
energy, which itself depends on the electron density. In this way 
the Sommerfeld $\gamma$ depends on the electron density. 

The density of states also reflects the electron band structure.
For example, the saddle points giving a high density of states are 
quite sensitive to atomic spacing. Moreover, in ionic crystals of
high-$T_{\rm c}$ superconductors the charge transfer between 
sublattices depends on the lattice deformation. The Sommerfeld 
$\gamma$ thus depends on the lattice deformation via mechanisms which 
are distinct from changes of the electron density.

The critical temperature is an even more complex quantity. 
For simplicity we express it within the BCS approximation
$T_{\rm c}=0.85\,\Theta_{\rm D}{\rm exp}(-1/N_0V)$. Apparently, 
$T_{\rm c}$ depends on the electron density and the deformation 
via the density of states. Beside this, there are additional contributions 
via the interaction potential $V$ and the Debye
temperature  $\Theta_{\rm D}$. For example, a compression of the 
lattice increases the mass density which results in a slower velocity 
of sound. This reduces the Debye temperature $\Theta_{\rm D}$
leading in some superconductors to a decrease of $T_{\rm c}$ under 
an applied pressure.\cite{JS58}

The above mentioned mechanisms of the density dependence work
in pure materials. Let us mention a mecha\-nism specific for dirty
superconductors. In metals doped by paramagnetic impurities the
dominant pressure dependence of $T_{\rm c}$ results from
the electron density dependence of the magnetic scattering relaxation 
time.\cite{Smith66}

\subsection{Plan of the paper}
In section~\ref{secFE} we introduce the free energy which combines
the condensation energy of Ginzburg and Landau (GL), the energy of
electric and magnetic fields, and the deformation energy. In 
section~\ref{secLVC} we present the set of equations derived 
from the Lagrange variational principle. In sections~\ref{secBP} and~\ref{secSE} 
we focus on the electrostatic potential and the strain, respectively.
In section~\ref{secEFE} we write down an effective free energy for
deformable superconductors, and section~\ref{secFC} is the summary.

\section{Free energy}\label{secFE}
We start from the free energy and employ the Lagrange 
variational principle to derive all stability conditions.
Following Ginzburg and Landau (GL), the free energy of the 
superconducting state
\begin{align}
f_{\rm s}&=f_{\rm n}+\alpha|\psi|^2+{1\over 2}\beta\,|\psi|^4
+{1\over 2m^*}\left|\left(-i\hbar\nabla-e^*{\bf A}\right)\psi\right|^2
\label{r1}
\end{align}
is defined as a non-local bi-quadratic function of the
GL function $\psi$. The non-local term has a form of the
kinetic energy with the Copper pair mass $m^*$ and charge
$e^*=2e$. 

The GL free energy is added to the free energy of the normal state
\begin{equation}
f_{\rm n}=f_0+e\varphi(n-n_{\rm lat})-
{1\over 2}\epsilon\left|\nabla\varphi\right|^2+
{1\over 2\mu_0}
\left|\nabla\times{\bf A}\right|^2.
\label{r2}
\end{equation}
The normal free energy covers the magnetic energy (last term),
the electrostatic energy (second and third terms) and the 
local free energy $f_0$.

The local free energy $f_0$ is a function of the electron density 
$n$, deviations of atomic positions $\bf u$ and temperature $T$
\begin{equation}
f_0=f_0(n,{\bf u},T).
\label{r3}
\end{equation}
According to the theory of elasticity,\cite{LL75} the free energy 
does not depend directly on the vector $\bf u$ but only on its 
derivatives expressed via the strain tensor
\begin{equation}
u_{ij}={1\over 2}\left({\partial u_i\over\partial r_j}+
{\partial u_j\over\partial r_i}\right).
\label{r4}
\end{equation}

There is no term explicitly attributed to the interaction
between the deformation $\bf u$ and the superconducting condensate,
however this interaction has a number of hidden contributions.
First, it is mediated by the electrostatic force. Second, the 
GL parameters $\alpha,\beta, m^*$ depend on the density of electrons
and on the deformation $\bf u$. 

For simplicity we will assume that the GL parameters $\alpha$, $\beta$ 
and $m^*$ depend on the strain exclusively via the crystal density 
$n_{\rm lat}$. In other words we neglect effects of shear deformations 
which break the isotropy of the system. The reader interested in the 
anisotropic interaction is referred to papers by Miranovi{\'c} 
{\em et al}\cite{MDK95} or Cano {\em et al}\cite{CLM03}.

In general, the lattice charge density in deformed crystal is a 
non-trivial problem since $n_{\rm lat}$ describes the ionic charge
and the deformation can change ionicity. We
will assume that the ionicity is constant so that the
charge is given by the divergence of atomic shifts
\begin{equation}
n_{\rm lat}=n_0\left(1-(\nabla\cdot{\bf u})\right)=n_0\left(
1-u_{11}-u_{22}-u_{33}\right).
\label{r5}
\end{equation}

Now all components of the free energy and material dependencies 
are specified. It remains to derive the equations for the individual
fields. To this end we employ the Lagrange variational principle.

\section{Lagrange variational conditions}\label{secLVC}
The free energy $f_{\rm s}$ depends on the following independent 
variable fields: 
the vector potential $\bf A$,
the complex GL function $\psi$, 
the electrostatic potential $\varphi$,
the electron density $n$, and
the vector of atomic shifts $\bf u$.
The corresponding variations 
are well established,\cite{Tinkham,W96,LKMBY07} therefore we 
present the resulting equations without derivations. We note that 
the vector and scalar potentials are in the Coulomb gauge, 
$(\nabla\cdot{\bf A})=0$.

The $\bf A$-variation yields the Ampere law
\begin{equation}
\nabla^2{\bf A}=-\mu_0{e^*\over m^*}{\rm Re}~
\bar\psi\left(-i\hbar\nabla-e^*{\bf A}\right)\psi.
\label{r6}
\end{equation}
The $\bar\psi$-variation results in the GL equation
\begin{equation}
\left(-i\hbar\nabla-e^*{\bf A}\right){1\over 2m^*}
\left(-i\hbar\nabla-e^*{\bf A}\right)\psi+
\alpha\psi+\beta|\psi|^2\psi=0.
\label{r7}
\end{equation}
The less usual form of the kinetic energy is a hermitian operator
also for an inhomogeneous mass $m^*$. The dependencies of the 
material parameters $\alpha$, $\beta$ and $m^*$ on 
the lattice deformation $\bf u$ and the electron density $n$ are 
rather weak but essential for specific problems like the vortex 
pinning by the strain around dislocations or for the effect of the 
elastic energy on the arrangement of vortices. 

The $\varphi$-variation recovers the Poisson equation
\begin{equation}
-\epsilon\nabla^2\varphi=e(n-n_{\rm lat}).
\label{r8}
\end{equation}
In deriving this equation we have neglected the 
ionic contribution to the dielectric function $\epsilon$, i.e., terms
proportional to $\partial\epsilon/\partial n_{\rm lat}$. 

The $n$-variation furnishes us with the electrostatic 
potential known as the Bernoulli potential
\begin{eqnarray}
e\varphi&=&-{\partial f_0\over\partial n}-
{\partial\alpha\over\partial n}|\psi|^2-
{1\over 2}{\partial\beta\over\partial n}|\psi|^4
\nonumber\\
&+&\bar\psi\left(-i\hbar\nabla-e^*{\bf A}\right)
{1\over 2m^{*}}{\partial\ln m^*\over\partial n}
\bar\psi\left(-i\hbar\nabla-e^*{\bf A}\right)\psi.
\nonumber\\
\label{r9}
\end{eqnarray}
The density derivative
of the local free energy $f_0$ is non-trivial only if the
system is perturbed from the homogenous state. To make 
$\partial f_0/\partial n$
transparent, it is necessary to expand it in perturbations. 
This rearrangement is accomplished in section~\ref{secBP}.

The $\bf u$-variation gives the strain equation 
\begin{align}
\nabla_j&{\partial f_0\over\partial u_{ij}}=
-\nabla_j{\partial\alpha\over\partial u_{ij}}|\psi|^2-
{1\over 2}\nabla_j{\partial\beta\over\partial u_{ij}}|\psi|^4+
\nabla_je\varphi{\partial n_{\rm lat}\over\partial u_{ij}}
\nonumber\\
&+\nabla_j
\bar\psi\left(-i\hbar\nabla-e^*{\bf A}\right)
{1\over 2m^{*}}{\partial\ln m^*\over\partial u_{ij}}
\bar\psi\left(-i\hbar\nabla-e^*{\bf A}\right)\psi.
\nonumber\\
\label{r10}
\end{align}
We use the Einstein summation rule for doubled indices, e.g.,
$r_{j}h_{jm}\equiv\sum_{j=1}^3r_{j}h_{jm}$.
The strain equation (\ref{r10}) includes terms which are so far rather
symbolic. In section~\ref{secSE} we express all of them in terms 
of elastic moduli and forces on the crystal lattice.

\section{Bernoulli potential}\label{secBP}
We start with a rearrangement of the Bernoulli potential (\ref{r9}).
The first derivative of the local free energy with respect to
the electron density is the Fermi energy
\begin{equation}
{\partial f_0\over\partial n}=E_{\rm F}.
\label{r11}
\end{equation}
The Fermi energy itself depends on the electron density via the 
Fermi-Dirac statistics and the exchange-correlation 
potential.\cite{Drez90} 
Besides, it depends on the lattice deformation via the density 
of states. Setting the Fermi energy of unperturbed system to zero, 
to the linear order in perturbations it reads
\begin{equation}
E_{\rm F}=
{\partial E_{\rm F}\over\partial n}\delta n+
{\partial E_{\rm F}\over\partial u_{ij}}u_{ij}.
\label{r11ad}
\end{equation}

The Fermi energy $E_{\rm F}$ depends on the lattice deformation 
via changes of the electron band structure. Within the isotropic 
approximation we assume that it is proportional to the perturbation of
the lattice density
\begin{equation}
{\partial E_{\rm F}\over\partial u_{ij}}=
{\partial E_{\rm F}\over\partial n_{\rm lat}}
{\partial n_{\rm lat}\over\partial u_{ij}}=-
{\partial E_{\rm F}\over\partial n_{\rm lat}}n_0
\delta_{ji}.
\label{r12}
\end{equation}
Using the approximation (\ref{r12}) in the relation (\ref{r11}) 
we obtain [$u_{ii}\equiv u_{11}+u_{22}+u_{33}$]
\begin{equation}
{\partial f_0\over\partial n}=
{\partial E_{\rm F}\over\partial n}\delta n-
{\partial E_{\rm F}\over\partial n_{\rm lat}}n_0u_{ii}.
\label{r13}
\end{equation}
The first term represents the Thomas-Fermi screening, the second
one results from the charge inhomogeneity of the deformed ionic 
lattice.

\subsection{Thomas-Fermi Screening}\label{subsecTFS}
Now we express the change of the Fermi energy in terms of the
electrostatic potential $\varphi$. To this end we use the Poisson 
equation (\ref{r8}) in the form
\begin{equation}
-\epsilon\nabla^2\varphi=e(\delta n+n_0u_{ii}).
\label{r14}
\end{equation}
Substituting $\delta n$ from equation (\ref{r14}) in the Fermi energy
(\ref{r13}) we arrive at
\begin{equation}
{\partial f_0\over\partial n}=-
{\partial E_{\rm F}\over\partial n}{\epsilon\over e}
\nabla^2\varphi-
\left({\partial E_{\rm F}\over\partial n}+
{\partial E_{\rm F}\over\partial n_{\rm lat}}\right)
n_0u_{ii}.
\label{r15}
\end{equation}

The first term on the right hand side can be expressed via
the Thomas-Fermi screening length
\begin{equation}
{\partial E_{\rm F}\over\partial n}{\epsilon\over e^2}=
\lambda_{\rm TF}^2.
\label{r16}
\end{equation}
The Bernoulli potential (\ref{r9}) now reads
\begin{align}
e\varphi&-\lambda_{\rm TF}^2\nabla^2e\varphi=
\left({\partial E_{\rm F}\over\partial n}+
{\partial E_{\rm F}\over\partial n_{\rm lat}}\right)n_0u_{ii}
\nonumber\\
&~~~~~~~~~~~~~~~~-
{\partial\alpha\over\partial n}|\psi|^2-
{1\over 2}{\partial\beta\over\partial n}|\psi|^4
\nonumber\\
&+\bar\psi\left(-i\hbar\nabla-e^*{\bf A}\right)
{1\over 2m^{*}}{\partial\ln m^*\over\partial n}
\left(-i\hbar\nabla-e^*{\bf A}\right)\psi.
\nonumber\\
\label{r17}
\end{align}

The electrostatic potential $\varphi$ resulting from equation 
(\ref{r17}) has two characteristic components, the free
and the enforced one. The free solution is non-zero only near 
the surface decaying into the bulk on the Thomas-Fermi screening 
length $\lambda_{\rm TF}$. This solution is determined by a 
surface condition. We note that the free solution plays an 
important role in the surface dipole.\cite{LKMBY07} Here we 
focus on the bulk properties, therefore we ignore the free 
solution.

Second, there is an electrostatic potential enforced by 
inhomogeneities in the superconducting density $|\psi|^2$ 
and the lattice deformation as given by the right hand 
side of equation (\ref{r17}). We keep the name Bernoulli
potential for this component. 

Two simplifications of the Bernoulli potential are at
hand. First, we can neglect $\lambda_{\rm TF}^2\nabla^2e\varphi$.
This is because gradients of the GL function and the corresponding
potential are on the scale of the GL coherence length or 
the London penetration which are both much larger than the 
Thomas-Fermi screening length $\lambda_{\rm TF}$. Second,
the logarithmic derivative of the Cooper pair mass $m^*$
is a small quantity and we can neglect its gradient.
Therefore
\begin{align}
e\varphi&=
\left({\partial E_{\rm F}\over\partial n}+
{\partial E_{\rm F}\over\partial n_{\rm lat}}\right)n_0u_{ii}-
{\partial\alpha\over\partial n}|\psi|^2-
{1\over 2}{\partial\beta\over\partial n}|\psi|^4
\nonumber\\
&+{\partial\ln m^*\over\partial n}
\bar\psi\left(-i\hbar\nabla-e^*{\bf A}\right)
{1\over 2m^{*}}
\left(-i\hbar\nabla-e^*{\bf A}\right)\psi.
\nonumber\\
\label{r17a}
\end{align}

Note that neglecting the term $\nabla^2\varphi$ in the Poisson
equation (\ref{r14}) implies the quasi-neutral approximation
$n=n_{\rm lat}$. In this sense we can work with the non-zero 
electrostatic potential (\ref{r17a}) while using the local 
charge neutrality for perturbations of material parameters.

The Bernoulli potential (\ref{r17a}) extends previous 
results\cite{LKMB01,LKMBY07} having two additional contribution. 
First, the charge of the deformed ion lattice is represented by
the term $\propto n_0u_{ii}=n_0(\nabla\cdot{\bf u})$. Second, the
effect of the charge perturbation on the Cooper pair mass $m^*$
is included.

\subsection{From non-local to non-linear corrections}\label{subsecNNC}

The GL equation (\ref{r7}) multiplied by the conjugate GL 
function $\bar\psi$
\begin{equation}
\bar\psi\left(-i\hbar\nabla-e^*{\bf A}\right){1\over 2m^*}
\left(-i\hbar\nabla-e^*{\bf A}\right)\psi=-
\alpha|\psi|^2-\beta|\psi|^4,
\label{r18}
\end{equation}
couples the non-local term on the left hand side with the 
non-linear one $\beta|\psi|^4$. This gives us the freedom to
make the Bernoulli potential either a linear or a local function
of the superconducting density $|\psi|^2$. We prefer the 
local but non-linear form,
\begin{eqnarray}
e\varphi&=&
\left({\partial E_{\rm F}\over\partial n}+
{\partial E_{\rm F}\over\partial n_{\rm lat}}\right)n_0u_{ii}
\nonumber\\
&-&
\left({\partial \alpha\over\partial n}+\alpha
{\partial\ln m^*\over\partial n}\right)|\psi|^2
\nonumber\\
&-&
{1\over 2}\left({\partial \beta\over\partial n}+2\beta
{\partial\ln m^*\over\partial n}\right)|\psi|^4.
\label{r19ad}
\end{eqnarray}

Apparently, there are a number of possible additional 
rearrangements of the Bernoulli potential. Since we
study the interaction between the superconducting condensate 
and the lattice deformation mediated by the Bernoulli potential,
the form (\ref{r19ad}) is optimal as it is expressed in terms 
of $u_{ii}$ and $|\psi|^2$.

\section{Strain equation}\label{secSE}
The strain equation (\ref{r10}) is rather involved as it
contains gradients of derivatives with respect to tensor
components of the strain. The major simplification follows
from the assumption that all material parameters related
to the superconducting phase depend on the strain exclusively 
via the lattice density $n_{\rm lat}$, i.e., 
\begin{align}
{\partial \alpha\over\partial u_{ij}}=-
{\partial \alpha\over\partial n_{\rm lat}}n_0\delta_{ji}
\label{r28}
\end{align}
and similar for $\beta$ and $m^*$. 
Within this isotropic approximation the strain equation can be 
rearranged in a manner which in many steps parallels the treatment 
of the Fermi energy in the previous section. 

\subsection{Stress}\label{secS}
The stress tensor has a general form of
\begin{equation}
p_{ji}=\Lambda_{jilk}u_{kl}.
\label{r35}
\end{equation}
The moduli matrix $\Lambda$ has 81 elements, but only 27
of them are independent.\cite{LL75}

Now we express the moduli tensor $\Lambda$ in terms of the 
free energy $f$. We start with the strain-derivative of the
local free energy
\begin{equation}
{\partial f_0\over\partial u_{ij}}=
{\partial^2 f_0\over\partial u_{ij}\partial u_{kl}}u_{kl}+
{\partial^2 f_0\over\partial u_{ij}\partial n}\delta n,
\label{r20}
\end{equation}
which we have expanded in perturbations.
The second term of expansion (\ref{r20}) we can express with the help of 
the already specified strain derivative of the Fermi energy
\begin{equation}
{\partial^2 f_0\over\partial u_{ij}\partial n}=
{\partial E_{\rm F}\over\partial u_{ij}}=-
{\partial E_{\rm F}\over\partial n_{\rm lat}}n_0\delta_{ji}.
\label{r21}
\end{equation}
Finally we use the Poisson equation (\ref{r14}) to eliminate the
perturbation of the electron density $\delta n$ from the stress
\begin{equation}
{\partial f_0\over\partial u_{ij}}=
{\partial^2 f_0\over\partial u_{ij}\partial u_{kl}}u_{kl}+
\delta_{ji}u_{kk}
{\partial E_{\rm F}\over\partial n_{\rm lat}}n_0^2.
\label{r23}
\end{equation}
We have neglected the term $\delta_{ji}{\partial E_{\rm F}\over
\partial n_{\rm lat}}n_0{\epsilon_0\over e}\nabla^2\varphi$, 
because it is proportional to $\lambda_{\rm TF}^2\nabla^2\varphi$.

An additional contribution to the stress results from the Coulomb
interaction of the ionic lattice with itself. To make it explicit,
we have to rearrange the electrostatic term of the strain equation 
(\ref{r10}) with the help of the Bernoulli potential (\ref{r19ad})
\begin{eqnarray}
e\varphi{\partial n_{\rm lat}\over\partial u_{ij}}&=&-
\delta_{ji}n_0e\varphi
\nonumber\\
&=&-
\delta_{ji}n_0\left({\partial E_{\rm F}\over\partial n}+
{\partial E_{\rm F}\over\partial n_{\rm lat}}\right)n_0u_{ii}
\nonumber\\
&&+
\delta_{ji}n_0\left({\partial \alpha\over\partial n}+\alpha
{\partial\ln m^*\over\partial n}\right)|\psi|^2
\nonumber\\
&&+
\delta_{ji}n_0{1\over 2}\left({\partial \beta\over\partial n}+2\beta
{\partial\ln m^*\over\partial n}\right)|\psi|^4.
\label{r23a}
\end{eqnarray}
The first term represents the Coulomb interaction of the lattice
with itself. Other terms represent the interaction of the lattice
with the superconducting condensate.

The stress tensor collects all contributions to the strain
equation (\ref{r10}) which are linear in the strain $u$.
The moduli matrix thus reads
\begin{equation}
\Lambda_{jilk}=
{\partial^2 f_0\over\partial u_{ij}
\partial u_{kl}}+\delta_{ji}\delta_{lk}
\left(2{\partial E_{\rm F}\over\partial n_{\rm lat}}+
{\partial E_{\rm F}\over\partial n}\right)n_0^2.
\label{r32}
\end{equation}
Here the second term arises from the increase of the electron
liquid energy under a volume compression. 

\subsection{Deforming force}\label{secDF}

In terms of the stress (\ref{r35}) the strain equation 
(\ref{r10}) reads
\begin{equation}
\nabla_jp_{ji}=F_i,
\label{r24p}
\end{equation}
where 
\begin{eqnarray}
&&F_i=\nabla_j\left(
\delta_{ji}n_0\left({\partial \alpha\over\partial n}+\alpha
{\partial\ln m^*\over\partial n}\right)-
{\partial \alpha\over\partial u_{ij}}\right)|\psi|^2
\nonumber\\
&&~~~+\nabla_j\left(
\delta_{ji}n_0{1\over 2}\left({\partial \beta\over\partial n}+2\beta
{\partial\ln m^*\over\partial n}\right)-{1\over 2}
{\partial \beta\over\partial u_{ij}}\right)|\psi|^4
\nonumber\\
&&+\nabla_j{\partial\ln m^*\over\partial u_{ij}}
\bar\psi\left(-i\hbar\nabla-e^*{\bf A}\right){1\over 2m^{*}}
\bar\psi\left(-i\hbar\nabla-e^*{\bf A}\right)\psi
\nonumber\\
\label{r24}
\end{eqnarray}
is the force (per unit volume) deforming the crystal.
We have neglected the gradient of $\partial\ln m^*/
\partial u_{ij}$. 

In the isotropic approximations (\ref{r28}) the deforming force 
(\ref{r24}) simplifies to a gradient 
\begin{equation}
F_i=-n_0\nabla_i U
\label{r24a}
\end{equation}
of the effective potential
\begin{eqnarray}
&&U=-\left({\partial \alpha\over\partial n}+
{\partial \alpha\over\partial n_{\rm lat}}+
\alpha{\partial\ln m^*\over\partial n}+
\alpha{\partial\ln m^*\over\partial n_{\rm lat}}
\right)|\psi|^2
\nonumber\\
&&~~-{1\over 2}
\left({\partial \beta\over\partial n}+
{\partial \beta\over\partial n_{\rm lat}}
+2\beta{\partial\ln m^*\over\partial n}
+2\beta{\partial\ln m^*\over\partial n_{\rm lat}}
\right)|\psi|^4.
\nonumber\\
\label{r24c}
\end{eqnarray}
We have used (\ref{r18}) to replace the non-local 
term of (\ref{r24}) by the non-linear one.

As one can see, the force is given by the bi-quadratic 
effective potential with two material parameters 
\begin{align}
a&={\partial \alpha\over\partial n_{\rm lat}}+
{\partial \alpha\over\partial n}
+\alpha{\partial\ln m^*\over\partial n_{\rm lat}}
+\alpha{\partial\ln m^*\over\partial n},
\nonumber\\
b&={\partial \beta\over\partial n_{\rm lat}}+
{\partial \beta\over\partial n}
+2\beta{\partial\ln m^*\over\partial n_{\rm lat}}
+2\beta{\partial\ln m^*\over\partial n}.
\label{r33}
\end{align}
In both terms the derivatives enter in the
same way as if one takes the volume or density derivative
assuming the strict local charge neutrality. 

\subsection{Isotropic model}\label{secIC}

The simplest and mostly used isotropic model uses only two 
elastic moduli. The bulk modulus $K$ measures changes
of the specific volume and shear modulus $\mu$ is the only 
coefficient of all volume-keeping deformations. For the 
isotropic system the strain equation (\ref{r24p}) 
simplifies to\cite{LL75}
\begin{equation}    
\left(K+{4\over 3}\mu\right)\nabla(\nabla.{\bf u})-
\mu\nabla\times\nabla\times{\bf u}={\bf F},
\label{r36}
\end{equation}
where the force acting on a unit volume of the lattice is given 
by the gradient as
\begin{equation}
{\bf F}=a\nabla|\psi|^2+{1\over 2}b\nabla|\psi|^4.
\label{r37}
\end{equation}
Together with (\ref{r33}) this is our final result for the strain equation. 

\section{Effective free energy}\label{secEFE}
For studies of the lattice deformations it is not necessary to 
evaluate the electrostatic potential. In this case one can use 
a simplified free energy
\begin{align}
f_{\rm s}'&=\alpha|\psi|^2+{1\over 2}\beta|\psi|^4+{1\over 2m^*}
\left|(-i\hbar\nabla-e^*{\bf A})\psi\right|^2
\nonumber\\ 
&
+{1\over 2\mu_0}
\left|\nabla\times{\bf A}\right|^2+
{1\over 2}\Lambda_{ijkl}u_{ij}u_{kl}
\nonumber\\   
&
-a\,u_{ii}|\psi|^2-{1\over 2}b\,u_{ii}|\psi|^4.
\label{r38}
\end{align}
This free energy depends on the vector potential $\bf A$,
the GL function $\psi$, and the displacement $\bf u$. All
material parameters $\alpha$, $\beta$, $m^*$, $\Lambda$, 
$a$ and $b$
are now constant in space and do not undergo variations.
 
By the Lagrange variation of the free energy 
$f_{\rm s}'$ with respect to the vector potential $\bf A$
one recovers the Ampere law  (\ref{r6}). The variation of
$f_{\rm s}'$ with respect to the displacement $\bf u$ yields 
the strain equation (\ref{r24p}) with the force (\ref{r37}).

The effective free energy $f_{\rm s}'$ is not exactly 
equivalent to the full free energy $f_{\rm s}$, however.
By the variation of $f_{\rm s}'$ with respect to the GL 
function $\bar\psi$ one arrives at the GL equation 
\begin{equation}
{1\over 2m^*}\left(-i\hbar\nabla-e^*{\bf A}\right)^2\psi+
(\alpha-a\,u_{ii})\psi+(\beta-b\,u_{ii})|\psi|^2\psi=0.
\label{r39}
\end{equation}
Unlike the full GL equation (\ref{r7}), here the strain
effect on the Copper pair mass $m^*$ is absent. It is mimicked by 
the $m^*$-part of the strain effect on the effective potential, 
see $a$ and $b$ as given by equations (\ref{r33}). 

\section{Summary}\label{secFC}
Starting from the free energy of the GL type we have derived 
the force which deforms the crystal lattice in the presence 
of the inhomogeneous superconducting condensate. Neglecting 
terms proportional to the square of the small Thomas-Fermi 
screening length, we have rearranged the deforming force into 
the gradient of the bi-quadratic function of the GL function. 

Although we took into account perturbations of the charge
neutrality and included the electrostatic potential, our
result has confirmed that the assumption of the strict local 
charge neutrality can be applied for the evaluation of the force
deforming the lattice. 

Based on our results, we have proposed an effective free 
energy which is simpler in being independent of the 
electrostatic potential and the density of normal elect\-rons. 
Moreover, all its field variables are explicit so that there
are no hidden interaction mechanisms. In particular, it has no 
strain effect on the Copper pair mass $m^*$. Contributions
of these eliminated variables and dependencies are covered
by the effective local but non-linear interaction.

\acknowledgements
This work was supported by research plans MSM 0021620834 and 
No. AVOZ10100521, by grants GA\v{C}R 202/07/0597, 202/08/0326 
and 202/06/0040 and GAAV 100100712 and IAA1010404, by PPP project 
of DAAD, by DFG Priority Program 1157 via GE1202/06 and the BMBF 
and by European ESF program NES.

\bibliography{bose,delay2,delay3,gdr,genn,chaos,kmsr,kmsr1,kmsr2,kmsr3,kmsr4,kmsr5,kmsr6,kmsr7,micha,refer,sem1,sem2,sem3,short,spin,spin1,solid,deform}

\end{document}